\documentclass{ws-procs9x6}
\newcommand{\RTargmin}{\mathop{\rm arg~min}\limits}
\usepackage{braket}

\begin{document}
 
\title{
A METHOD TO CHANGE PHASE TRANSITION NATURE\\
-- TOWARD ANNEALING METHODS --
}

\author{RYO TAMURA}

\address{
International Center for Young Scientists, National Institute for Materials Science,\\
1-2-1, Sengen, Tsukuba-shi, Ibaraki, 305-0047, Japan\\
E-mail: tamura.ryo@nims.go.jp
}

\author{SHU TANAKA}

\address{
Department of Chemistry, University of Tokyo,\\
7-3-1, Hongo, Bunkyo-ku, Tokyo, 113-0033, Japan\\
E-mail: shu-t@chem.s.u-tokyo.ac.jp
}

\begin{abstract}
In this paper, we review a way to change nature of phase transition with annealing methods in mind.
Annealing methods are regarded as a general technique to solve optimization problems efficiently.
In annealing methods, we introduce a controllable parameter which represents a kind of fluctuation and decrease the parameter gradually.
Annealing methods face with a difficulty when a phase transition point exists during the protocol.
Then, it is important to develop a method to avoid the phase transition by introducing a new type of fluctuation.
By taking the Potts model for instance, we review a way to change the phase transition nature.
Although the method described in this paper does not succeed to avoid the phase transition,
we believe that the concept of the method will be useful for optimization problems.
\end{abstract}


\bodymatter

\section{Introduction}
\label{RTsec:introduction}

Development of methods to solve optimization problems has been definitely a central issue in science.
Optimization problems are spread in a wide area of science such as mathematics, physics, chemistry, biology, and information science\cite{RTStrang-1986,RTMiller-1993,RTMartin-2001,RTHartmann-2005}.
Moreover, since optimization problems relate to phenomena in real world and daily life, development of useful optimization methods contributes to growth of industry.
Typical examples of optimization problems are designing of transportation system and that of integrated circuit.
Optimization problems are expressed by mathematically well-defined models.
In terms of mathematics, the goal of optimization problems is to find ${\bf x}^* := \RTargmin_{\bf x}f({\bf x})$, where $f({\bf x})$ is a real-valued function and called cost function. 
Here, ${\bf x}^*$ is referred to as the best solution.
When the cost function $f({\bf x})$ is defined by a simple form, we can easily differentiate $f({\bf x})$ and immediately obtain ${\bf x}^*$.
In general, however, since $f({\bf x})$ is a complicated function in optimization problems, 
to obtain ${\bf x}^*$ directly is difficult.
Then, we should develop a method to obtain the best solution of optimization problems.
There are many types of optimization problems.
Depending on individual types of optimization problems, many efficient but specialized algorithms have been developed mainly in information science\cite{RTMor-1987}.

As mentioned above, to solve optimization problems corresponds to find the state ${\bf x}^*$ which minimizes the cost function $f({\bf x})$.
As will be shown later, we can relate a cost function to a Hamiltonian of spin system in most cases.
In terms of physics, to solve optimization problems is to find the ground state of the corresponding Hamiltonian.
Then, to obtain the best solution of optimization problems, we can use methods developed in physics.
A generic algorithm was proposed in the context of physics, which imitates natural phenomena.
The most famous one is called simulated annealing\cite{RTKirkpatrick-1983,RTKirkpatrick-1984,RTDavis-1987,RTLaarhoven-1987,RTSzu-1987,RTAarts-1989,RTIngber-1993,RTGoffe-1994,RTTsuzuki-2012}.
``Annealing'' is a technical terminology in materials science.
Annealing is a gradual cooling process of metal alloys and glassy materials to remove stress and defects after these materials are synthesized.
The simulated annealing imitates the annealing in computer simulation, which is the origin of the terminology.
In the simulated annealing, temperature is introduced into optimization problems as thermal fluctuation.
In principle, the best solution can be obtained by decreasing the temperature gradually\cite{RTGeman-1984}.
Since the simulated annealing is easy to implement, it has been often used in many optimization problems.

There is another typical fluctuation in physics -- quantum fluctuation.
Annealing method in which quantum fluctuation is controlled was also proposed.
This method is called quantum annealing\cite{RTFinnila-1994,RTKadowaki-1998a,RTBrooke-1999,RTFarhi-2001,RTSantoro-2002,RTDas-2005,RTBattaglia-2006,RTSantoro-2006,RTDas-2008,RTOhzeki-2011,RTFalco-2011,RTSuzuki-2012,RTBapst-2013}.
In the quantum annealing, a quantum field which represents quantum fluctuation effect is introduced into optimization problems, and we gradually decrease the quantum field.
In principle, the best solution of optimization problems can be obtained as well as the simulated annealing\cite{RTMorita-2006,RTMorita-2007}.
In fact, a quantum field in the quantum annealing plays a similar role with the temperature in the simulated annealing.
Since the quantum annealing is easy to implement as the simulated annealing, it has been expected to be an alternative method to the simulated annealing.
Efficiency of the quantum annealing has been demonstrated in respective optimization problems.

Annealing methods such as the simulated annealing and quantum annealing seem to be efficient in general.
However, there is a serious crisis in annealing methods.
It becomes difficult to obtain the best solution by these methods if a phase transition point exists in the process of annealing.
Then, it is indispensable to develop a way to avoid the phase transition in optimization problems.
In other words, we should discover an annealing process in which no phase transition point exists.

In order to control phase transition behavior on demand, we first should establish a microscopic mechanism to change nature of phase transition.
To achieve the issue, we focus on the Potts model which has been used for analysis of phase transition with discrete symmetry breaking\cite{RTPotts-1952,RTWu-1982}.
The Potts model is a fundamental model in statistical physics and a straightforward generalization of the Ising model.

In this paper, we review a method to change nature of phase transition toward annealing methods.
The rest of this paper is organized as follows.
In Sec.~2, we review optimization problems with discrete variables and show relation between optimization problems and discrete spin systems which are typical models in statistical physics.
In Sec.~3, we explain annealing methods which have developed in physics and have been used to solve optimization problems.
In Sec.~4, nature of phase transition is considered in a general way.
In Sec.~5, we review properties and phase transition behavior of the Potts model with invisible states.
In the Potts model with invisible states, the order of phase transition is changed by controlling the number of invisible states.
Section 6 is devoted to conclusion and future perspective.

\section{Optimization problems}
\label{RTsec:op}

Optimization problems relate to many real-world problems which are concerned with maximizing benefit or minimizing cost.
As stated in Sec.~\ref{RTsec:introduction}, to solve optimization problems is to find the best solution ${\bf x}^*:=\RTargmin_{{\bf x}}f({\bf x})$.
The cost function of most optimization problems with discrete variables can be represented by Hamiltonian of discrete spin systems such as the Ising model and its generalizations.

Here we explain how to express the traveling salesman problem using the Ising model.
The traveling salesman problem is a typical optimization problem with discrete variables\cite{RTDantzig-1954,RTLin-1965,RTLawler-1985,RTJohnson-1997,RTDorigo-1997,RTGutin-2007,RTApplegate-2007}.
In the traveling salesman problem, the complete lists of cities and distances between two cities are given.
Let $N$ and $\ell_{i,j}$ be the number of cities and the distance between the $i$-th and $j$-th cities ($1 \le i,j \le N$), respectively.
By definition, $\ell_{i,j} = \ell_{j,i}$.
The traveling salesman problem is to find the shortest path under the following two conditions.
The first one is that a traveller can pass through an individual city just one time.
The second one is that a traveller finally returns to the initial city.
In other words, the start point is the same as the end point.
The cost function of traveling salesman problem is the length of path, which is represented by
\begin{eqnarray}
 \label{RTeq:Ham_of_TSP1}
 {\cal H} = \sum_{a=1}^N \ell_{c_a,c_{a+1}},
\end{eqnarray}
where $c_a$ denotes the city where a traveller passes through at the $a$-th step.
Because of the second condition and $\ell_{i,j}=\ell_{j,i}$, we can choose the initial city arbitrary and $c_{N+1}=c_1$ should be satisfied.
Then, the traveling salesman problem is to find $\{c_a\}_{a=1}^N$ such that the cost function ${\cal H}$ has the minimum value.
To express the cost function using a Hamiltonian of a discrete spin model, we introduce a new variable $n_{i,a}(=0,1)$ which represents the state of the $i$-th city at the $a$-th step.
When a traveller passes through the $i$-th city at the $a$-th step, $n_{i,a}=1$ whereas $n_{i,a}=0$ when a traveller passes through other city at the $a$-th step.
The first condition of traveling salesman problem can be represented by 
\begin{eqnarray}
 \label{RTeq:cond_TSP1}
 \sum_{a=1}^N n_{i,a}=1,
  \qquad
  \forall i (=1,\cdots,N).
\end{eqnarray}
Obviously, since a traveller passes through only one city in a single step, the following condition should be satisfied:
\begin{eqnarray}
 \label{RTeq:cond_TSP2}
 \sum_{i=1}^N n_{i,a}=1,
  \qquad
  \forall a (=1,\cdots,N).
\end{eqnarray}
Then, the cost function given by Eq.~(\ref{RTeq:Ham_of_TSP1}) is rewritten by
\begin{eqnarray}
 {\cal H} = \sum_{a=1}^N \sum_{i,j} \ell_{i,j} n_{i,a} n_{j,a+1}.
\end{eqnarray}
This cost function can be expressed by the Ising variable as
\begin{eqnarray}
 {\cal H} = \frac{1}{4} \sum_{a=1}^N \sum_{i,j} \ell_{i,j} \sigma_{i,a}^z \sigma_{j,a+1}^z + {\rm const.},
  \qquad
  \sigma_{i,a}^z=\pm 1.
\end{eqnarray}
Here we used the correspondence between the variable $n_{i,a}$ and the Ising variable $\sigma_{i,a}^z$:
\begin{eqnarray}
 n_{i,a}=\frac{1}{2}\left( \sigma_{i,a}^z+1\right).
\end{eqnarray}
Then, the conditions given by Eqs.~(\ref{RTeq:cond_TSP1}) and (\ref{RTeq:cond_TSP2}) are rewritten by 
\begin{eqnarray}
 \label{RTeq:cond_TSP3}
 &&\sum_{a=1}^N \sigma_{i,a}^z = -N+2,
  \qquad
  \forall i(=1,\cdots,N),\\
 \label{RTeq:cond_TSP4}
 &&\sum_{i=1}^N \sigma_{i,a}^z = -N+2,
  \qquad
  \forall a(=1,\cdots,N).
\end{eqnarray}
We can represent the cost function of traveling salesman problem by the Hamiltonian of Ising model with inhomogeneous interactions on $N\times N$ spins.
The number of microscopic states of this system is ${\cal O}(2^{N^2})$.
When the number of cities $N$ is small, we can easily obtain the ground state by a brute force.
However, since the number of microscopic states exponentially increases with $N^2$, it is difficult to obtain the ground state of the system for large $N$.

Here, we focus on traveling salesman problem.
As mentioned above, the cost function of most optimization problems can be represented by Hamiltonian of discrete spin systems with inhomogeneous interactions as well as the traveling salesman problem.
Then, we often face with the same difficulty to solve optimization problems in general.
We should develop an intelligent method to obtain the ground state.
In information science, efficient but specialized algorithms have been developed to solve respective optimization problems.
On the contrary, a generic algorithm was proposed in terms of physics.
The most famous algorithm is simulated annealing which will be explained in the next section.

\section{Annealing methods}
\label{RTsec:annealing}

In order to solve optimization problems, a generic algorithm called simulated annealing was proposed in a physical context\cite{RTKirkpatrick-1983,RTKirkpatrick-1984}.
In the simulated annealing, we introduce the temperature into optimization problems.
Since cost function of most optimization problems can be expressed by Hamiltonian of discrete spin systems, the temperature in optimization problem is well-defined.
At high temperatures, all states are realized with almost the same probability.
In contrast, at zero temperature, the system should be the ground state.
Next we gradually decrease the temperature.
In principle, the best solution of optimization problems can be definitely obtained when we decrease the temperature slow enough, which was mathematically proved\cite{RTGeman-1984,RTAarts-1984}.
Any system is guaranteed to converge to the stable state in the limit of infinite time if the temperature is decreased in proportion to inverse of the logarithm of time or slower.
Since the simulated annealing is easy to implement, it has been adopted for many optimization problems.

After the proposal of the simulated annealing, an alternative method to the simulated annealing -- quantum annealing, was proposed\cite{RTKadowaki-1998a}.
As mentioned above, the simulated annealing can obtain the best solution of optimization problems by imitating thermal fluctuation effect.
In contrast, the quantum annealing uses quantum fluctuation effect which is another fluctuation in nature.
In the quantum annealing, we introduce a quantum field into optimization problems.
For example, if a cost function of an optimization problem is described by the Ising model, we often introduce the transverse magnetic field as a quantum field.
Next we decrease the quantum field gradually.
The protocol of the quantum annealing is the same as that of the simulated annealing.
Then, the quantum annealing is also easy to implement as well as the simulated annealing.
In addition, the best solution of optimization problems can be definitely obtained when we decrease the quantum field slow enough.
In Refs.~29 and 30,~
sufficient conditions for convergence of the quantum annealing were given. 
The strong ergodicity property is proved in three implementation methods of the quantum annealing for the transverse Ising model under a power decay of the transverse field. 
In Ref.~29,~
the authors considered the cases of the path-integral Monte Carlo method and the Green's function Monte Carlo method. 
In Ref.~30,~
the case of real-time Schr\"odinger equation was considered. 
The latter study is based on the idea reported in Ref.~41~
in which classical-quantum correspondence was proposed.
Recently, experimental demonstrations of the quantum annealing have been done\cite{RTJohnson-2011,RTOrtiz-2012,RTBoixo-2012,RTDickson-2013}.
In this way, the quantum annealing is expected to be an efficient algorithm to solve optimization problems as well as the simulated annealing\cite{RTKurihara-2009,RTSato-2009,RTTanaka-2011a,RTSato-2013,RTMortonak-2004,RTStella-2005,RTTitiloye-2011}.
In this section, we consider a mechanism of annealing methods from a viewpoint of statistical physics.

\subsection{Mechanism of simulated annealing}
\label{RTsec:SA}

In the simulated annealing, we gradually decrease the temperature $T$ and obtain the state when the temperature reaches to $T=0$.
To explain a mechanism of the simulated annealing, suppose we consider the Ising model on a square lattice with homogeneous ferromagnetic interaction.
The Hamiltonian of the system is given by
\begin{eqnarray}
 \label{RTeq:IsingHam}
 {\cal H} = - J \sum_{\langle i,j \rangle} \sigma_i^z \sigma_j^z,
  \qquad
  \sigma_i^z = \pm 1,
\end{eqnarray}
where $\langle i,j \rangle$ denotes the nearest-neighbor spin pairs on a square lattice.
The ground state of the system is completely ferromagnetic ordered state.
It is only necessary to consider the ferromagnetic Ising spin system to show thermal fluctuation effect though the ground state is trivial.
Hereafter, the Boltzmann constant is set to unity.

\begin{figure}[t]
\begin{center}
 \psfig{file=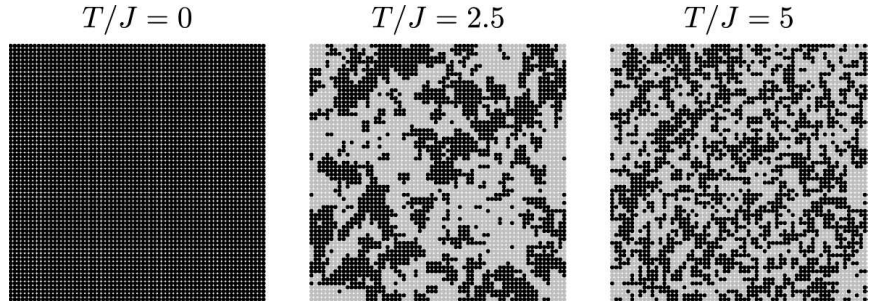,width=4.5in}
\end{center}
\caption{
Black and gray circles indicate $+1$ and $-1$ spins, respectively.
(Left panel) Perfectly ferromagnetic ordered state.
This is the ground state of the model given by Eq.~(\ref{RTeq:IsingHam}).
(Middle panel) A typical snapshot of spin configuration in equilibrium state at intermediate temperatures.
Short-range ferromagnetic correlation exists.
(Right panel) A typical snapshot of spin configuration in equilibrium state at high temperatures.
Random spin configuration appears.
}
\label{RTfig:Ising_equilibrium}
\end{figure}

First we consider the equilibrium state of the system on a $64 \times 64$ square lattice with periodic boundary condition.
At high temperatures, all spins are randomly oriented as shown in the right panel of Fig.~\ref{RTfig:Ising_equilibrium}.
As the temperature decreases, short-range ferromagnetic correlation grows, which is shown in the middle panel of Fig.~\ref{RTfig:Ising_equilibrium}.
At zero temperature, all spins are parallel, {\it i.e.}, completely ferromagnetic ordered state shown in the left panel of Fig.~\ref{RTfig:Ising_equilibrium} appears.
As described before, the ground state can be definitely obtained when we decrease the temperature slow enough\cite{RTGeman-1984}.
This is because the system stays close to the equilibrium at each temperature.

\begin{figure}[t]
 \psfig{file=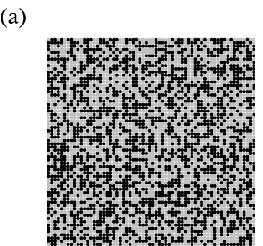,width=1.13in}
 \psfig{file=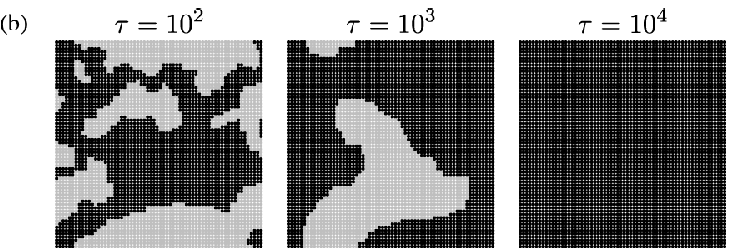,width=3.2in}
\begin{center}
 \psfig{file=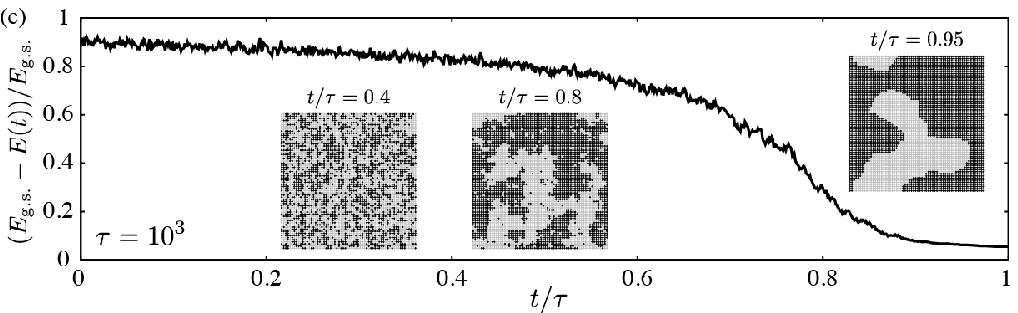,width=4.5in}
\end{center}
\begin{center}
 \psfig{file=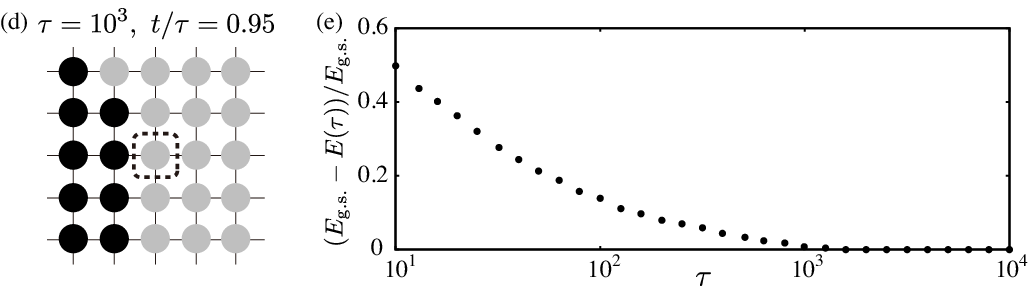,width=4.5in}
\end{center}
\caption{
(a) Initial state.
(b) The states at $t=\tau$ for $\tau=10^2,10^3,10^4$.
(c) Time development of the internal energy $E(t)$ for $\tau=10^3$.
Snapshots at several $t$'s are also shown.
(d) The enlarged view of domain wall.
(e) $\tau$-dependence of the internal energy $E(\tau)$.
}
\label{RTfig:Ising_cooling}
\end{figure}

In practice, however, we decrease the temperature with finite speed in the simulated annealing.
Then, it is important to consider dynamic nature of the Ising model.
There are many types of implementation methods of time evolution.
We now focus on the Monte Carlo method which is a stochastic method described in Appendix~A.
Here we adopt the heat-bath method as the transition probability (see Appendix~A).
We prepare a random spin configuration depicted in Fig.~\ref{RTfig:Ising_cooling}(a) as the initial state.
The initial temperature is set to $T_0/J=10$, which is much higher than the energy scale of magnetic interaction $J$.
Next we decrease the temperature with the following schedule:
\begin{eqnarray}
 \label{RTeq:schedule}
 T(t;\tau) = T_0\left( 1 - \frac{t}{\tau}\right),
 \qquad
 (0 \le t \le \tau),
\end{eqnarray}
where $\tau^{-1}$ is the sweeping speed.
At $t=\tau$, the temperature becomes zero.
Figure~\ref{RTfig:Ising_cooling}(b) shows snapshots of spin configuration at $t=\tau$ for various $\tau$'s.
Here we study the dynamics for $\tau=10^{3}$ in detail.
To consider the dynamics from a microscopic viewpoint, we calculate the internal energy $E(t)$ at $t$. 
To quantify the similarity between present state and the ground state, we consider the quantity: $(E_{\rm g.s.}-E(t))/E_{\rm g.s.}$ where $E_{\rm g.s.}$ is the internal energy of the ground state.
Figure~\ref{RTfig:Ising_cooling}(c) shows time development of the internal energy and snapshots at several $t$'s for $\tau=10^3$.
As shown in Fig.~\ref{RTfig:Ising_cooling}(c), the microscopic state does not almost change at all after $t/\tau=0.95$.
In other words, the state is trapped by the domain wall effect.
Here we estimate the probability to break domain walls.
The spin indicated by the dotted square in Fig.~\ref{RTfig:Ising_cooling}(d) flips with the probability:
\begin{eqnarray}
 p_{\rm flip} = \frac{{\rm e}^{-2\beta J}}{2\cosh (2\beta J)},
\end{eqnarray}
which is very small probability at low temperatures.
For example, at $T/J=0.5$ corresponding to $t/\tau=0.95$ where the snapshot began to almost stop, $p_{\rm flip}=0.00034$.
Then it is difficult to break domain wall once the domain wall forms.
In order to avoid the domain wall problem, we have to decrease the temperature as slow as possible.
Finally, we show $\tau$-dependence of the internal energy $E(\tau)$ obtained by the simulated annealing in Fig.~\ref{RTfig:Ising_cooling}(e).

\subsection{Mechanism of quantum annealing}
\label{RTsec:QA}

In the quantum annealing, we introduce a quantum field and gradually decrease the quantum field at zero temperature.
We obtain the state at zero quantum field as the final state.
In order to show a mechanism of the quantum annealing, we consider the Ising model with homogeneous ferromagnetic interaction as in the case of the simulated annealing.
When the cost function of optimization problem can be described by the Ising model, we often use the transverse field as the quantum field in the quantum annealing.
Then the total Hamiltonian is given by
\begin{eqnarray}
 \label{RTeq:TIM}
 \hat{\cal H} = -J \sum_{\langle i,j \rangle} \hat{\sigma}_i^z \hat{\sigma}_j^z - \Gamma \sum_{i=1}^N \hat{\sigma}_i^x,
\end{eqnarray}
where $\hat{\sigma}_i^\alpha$ denotes the $\alpha$-component of the Pauli matrix at the site $i$ ($\alpha=x,y,z$).
There are many types of implementation methods of the quantum annealing, 
for example, quantum Monte Carlo simulation, real-time dynamics, and time-dependent density matrix renormalization group (t-DMRG).\cite{RTSuzuki-2005,RTSuzuki-2007}
Here we focus on the quantum annealing using the real-time evolution which will be explained in Appendix B.

We consider $\Gamma$-dependence of eigenstates and eigenenergies.
The ground state depends on the magnitude of transverse field $\Gamma$.
When $\Gamma=0$, the ground state is completely ferromagnetic ordered state expressed as $\ket{\uparrow\uparrow\cdots\uparrow}$ or $\ket{\downarrow\downarrow\cdots\downarrow}$.
Here $\hat{\sigma}_i^z\ket{\uparrow}=\ket{\uparrow}$ and $\hat{\sigma}_i^z\ket{\downarrow}=-\ket{\downarrow}$.
In contrast, the ground state in the limit of $\Gamma \to \infty$ is represented by $\ket{\rightarrow\rightarrow\cdots\rightarrow}$, where $\hat{\sigma}_i^x\ket{\rightarrow}=\ket{\rightarrow} :=\frac{1}{\sqrt{2}}\left( \ket{\uparrow} + \ket{\downarrow}\right)$.
The purpose of the quantum annealing is to obtain the ground state at $\Gamma=0$.
In this case, the ground state at $\Gamma=0$ is trivial.
In general, however, it is difficult to obtain the ground state at $\Gamma=0$ of the Hamiltonian with inhomogeneous interactions for large $N$.
In contrast, the ground state of the Ising models at $\Gamma\to \infty$ is definitely a trivial state expressed as $\ket{\rightarrow\rightarrow\cdots\rightarrow}$.
Then, we can easily prepare the initial state and obtain the ground state at $\Gamma=0$ by just decreasing transverse field in the quantum annealing.

\begin{figure}[t]
\begin{center}
 \psfig{file=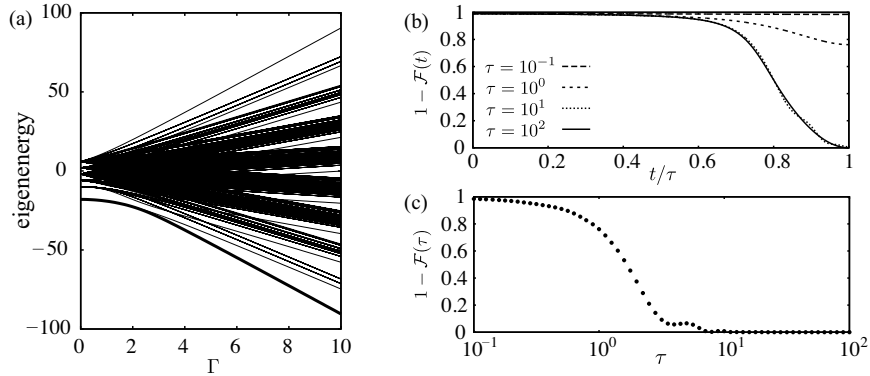,width=4.5in}
\end{center}
\caption{
(a) $\Gamma$-dependence of eigenenergies of the ferromagnetic Ising model on square lattice with $3\times 3$ sites.
The bold curve indicates the eigenenergy of the ground state.
(b) $1-\mathcal{F}(t)$ as a function of step $t$ for some sweeping speeds $\tau^{-1}$. 
(c) Sweeping speed $\tau^{-1}$-dependence of $1-\mathcal{F}(\tau)$.
}
\label{RTfig:Ising_QA}
\end{figure}

We calculate eigenenergies of the Hamiltonian given by Eq.~(\ref{RTeq:TIM}).
Figure \ref{RTfig:Ising_QA}(a) depicts eigenenergies of the model on $3 \times 3$ square lattice with periodic boundary condition.
The bold curve in Fig.~\ref{RTfig:Ising_QA}(a) displays $\Gamma$-dependence of eigenenergy of the ground state.
The curve is smoothly connected between the eigenenergy at large $\Gamma$'s and that at $\Gamma=0$. 
Thus, if we can prepare the ground state at finite $\Gamma$ as the initial state, we can definitely obtain the ground state at $\Gamma=0$ in the adiabatic limit.

In practice, however, we decrease the quantum field with finite speed.
Then, a nonadiabatic transition occurs during the protocol of the quantum annealing.
To show nonadiabatic transition effect, we demonstrate the quantum annealing.
The initial transverse field is set to be $\Gamma_0/J=10$, which is much larger than the scale of magnetic interaction $J$.
We prepare the ground state at $\Gamma_0/J$ as the initial state.
Next we decrease the transverse field with the following schedule:
\begin{eqnarray}
 \Gamma(t;\tau) = \Gamma_0 \left( 1 - \frac{t}{\tau}\right),
  \qquad
 (0 \le t \le \tau),
\end{eqnarray}
which is the same schedule as Eq.~(\ref{RTeq:schedule}).
As mentioned above, the ground state at $\Gamma=0$ is completely ferromagnetic ordered state.
The fidelity between the ground state and the state at $t$ obtained by the quantum annealing is calculated.
The fidelity is defined by 
\begin{eqnarray}
 {\cal F} (t):=\left|\braket{\psi(t)|\phi_{\rm g.s.}}\right|^2,
\end{eqnarray}
where $\ket{\psi(t)}$ is the wavefunction at $t$ obtained by the quantum annealing and $\ket{\phi_{\rm g.s.}}$ is the wavefunction of the ground state at $\Gamma=0$.
When the fidelity closes to 1, the present state obtained by the quantum annealing is similar with the ground state.
Figure~\ref{RTfig:Ising_QA}(b) shows the similarity between the present state and the ground state $1-\mathcal{F} (t)$ as a function of $t$ for some sweeping speeds $\tau^{-1}$.
As sweeping speed $\tau^{-1}$ increases, $1-\mathcal{F} (t)$ does not reach to zero.
Figure~\ref{RTfig:Ising_QA}(c) shows the sweeping speed $\tau^{-1}$-dependence of the fidelity at $t=\tau$.
As the sweeping speed $\tau^{-1}$ increases, $1-\mathcal{F} (\tau)$ increases, which comes from the nonadiabatic transition.
Then, in order to avoid the nonadiabatic transition, we have to decrease the quantum field as slow as possible.

\section{Phase transitions}

In Sec.~\ref{RTsec:annealing}, we reviewed mechanisms of the simulated annealing and that of the quantum annealing.
In both cases, we introduce a controllable parameter which represents some kind of fluctuation and gradually decrease the parameter.
We can prevent unpreferable transition to excited states by decreasing the fluctuation parameter as slow as possible.
Then, annealing methods such as the simulated annealing and quantum annealing seem to be versatile for optimization problems.
However, we face with difficulties which come from phase transition in annealing methods.
As shown above, cost function of most optimization problems with discrete variables can be represented by Hamiltonian of discrete spin systems.
According to statistical physics, discrete spin systems exhibit a phase transition in many cases.
If there is a transition point in the protocol of annealing methods, it becomes difficult to obtain the best solution.

Phase transitions are divided into two types according to singularity in physical quantities.
If the first-order derivative of the free energy is discontinuous, the transition is of the first order and called discontinuous phase transition or first-order phase transition.
When the second-order or higher-order derivative of the free energy is discontinuous or divergent, the transition is called continuous transition.
When the second-order derivative of the free energy is first discontinuous or divergent, the phase transition is called second-order phase transition.
In this section, we explain inherent problem in annealing methods when the system exhibits a phase transition.

\subsection{First-order phase transition}

When a first-order phase transition takes place in the protocol of annealing method, it is difficult to obtain the stable state because of existence of metastable states.
As stated above, the first derivative of free energy is discontinuous or divergent at the first-order phase transition point.
Typical example of first-order phase transition in nature is ice-water phase transition at $0^\circ$C under atmospheric pressure.
When we decrease the temperature rapidly under atmospheric pressure, water does not change into ice even below $0^\circ$C though the equilibrium state of H$_2$O is ice below $0^\circ$C.
This behavior is called supercooled phenomenon.
The supercooled phenomenon also appears in many magnetic and electronic compounds when we decrease the temperature rapidly.
In these materials, hysteresis curve of the physical quantities such as magnetization obtained by the first derivative of free energy is observed.
The hysteresis curve indicates the existence of metastable states.
Once the state is trapped in the metastable state, to reach the stable state is difficult.

The same situation happens in theoretical models in which a first-order phase transition occurs.
Typical examples of these models are the Blume-Emery-Griffiths model\cite{RTBlume-1971} and the Wajnflasz-Pick model\cite{RTWajnflasz-1971}.
The Hamiltonians of the Blume-Emery-Griffiths model and the Wajnflasz-Pick model are respectively represented by
\begin{eqnarray}
 \label{RTeq:BEGmodel}
 &&{\cal H}_{\rm BEG} = - J\sum_{\langle i,j \rangle} S_i S_j - J'\sum_{\langle i,j \rangle} S_i^2 S_j^2 - D\sum_{i} \left( S_i \right)^2,
  \quad
  S_i = \pm 1, 0,\\
 \label{RTeq:WPmodel}
 &&{\cal H}_{\rm WP} = - J \sum_{\langle i,j \rangle} S_i S_j,
  \qquad
  S_i = \underbrace{+1, \cdots, +1}_{g_+}, \underbrace{-1, \cdots, -1}_{g_-}.
\end{eqnarray}
The former Hamiltonian was proposed in order to explain the phase transition nature of ${}^3$He-${}^4$He mixture\cite{RTBlume-1971}.
The latter one can analyze the phase transition behavior of spin-crossover materials, and the number of $+1$ states and that of $-1$ states in the latter Hamiltonian are $g_+$ and $g_-$, respectively\cite{RTZimmermann-1983,RTBousseksou-1992,RTHauser1999,RTBoukheddaden-2000,RTMiyashita-2003,RTNishino-2003,RTTokoro-2006}.
Furthermore, we can transfer the Wajnflasz-Pick model into the following Hamiltonian at finite temperature $T$:
\begin{eqnarray}
 \label{RTeq:WPt}
 {\cal H}_{\rm WP} = -J \sum_{\langle i,j\rangle} \sigma_i^z \sigma_j^z 
  - (h-\frac{T}{2}\log \frac{g_+}{g_-}) \sum_i \sigma_i^z,
  \quad
  \sigma_i^z = +1, -1.
\end{eqnarray}
The bias of $g_+$ and $g_-$ induces the temperature-dependent chemical potential.

These models given by Eq.~(\ref{RTeq:BEGmodel}) and Eqs.~(\ref{RTeq:WPmodel}) and (\ref{RTeq:WPt}) are generalized Ising models and exhibit a thermal-induced first-order phase transition for a certain parameter region.
When we decrease the temperature rapidly, hysteresis curve appears in these models.
Since the cost function of optimization problems with discrete variables can be represented by Hamiltonian of discrete spin systems, a first-order phase transition sometimes occurs in optimization problems.
If the supercooled phenomenon occurs in optimization problems because of first-order phase transition, we cannot obtain the best solution definitely.
Then, in order to improve annealing method, we should avoid the first-order phase transition point in the protocol of annealing method.
In terms of the quantum annealing, difficulty in systems where a first-order phase transition appears can be explained as follows. The energy gap is likely to be exponentially small at the first-order phase transition point, which leads to exponential complexity\cite{RTAltshuler-2009,RTAmin-2009,RTAltshuler-2010}.
In Ref.~67,~
the authors concluded that the exact cover problem, which is a typical optimization problem, in the quantum annealing exhibits a first-order phase transition. 
Recently, the authors in Ref.~68~
studied antiferromagnetic fluctuation effect in the ferromagnetic $p$-spin model with transverse field with the quantum annealing in mind. 
Originally, the model exhibits a first-order phase transition. 
However, they found that we can make a path that avoids the first-order phase transition point.

\subsection{Second-order phase transition}

When a second-order phase transition occurs in optimization problems, we face with other type of difficulty caused by critical slowing down.
As described before, the first derivative of free energy is analytic but the second derivative of free energy is discontinuous or divergent at the second-order phase transition point.
Physical quantities can be described by power-law behavior near the second-order phase transition point.
Suppose we consider the magnetic system in which a second-order phase transition occurs at $T=T_{\rm c}$, where $T_{\rm c}$ is called critical temperature.
The specific heat $C$, magnetization $m$, and magnetic susceptibility $\chi$ near the critical point behave as
\begin{eqnarray}
 \label{RTeq:criticalexponents}
 C(T) \propto \left| T-T_{\rm c}\right|^{-\alpha},
  \quad
 m(T) \propto \left| T-T_{\rm c}\right|^\beta,
 \quad
 \chi(T) \propto \left| T-T_{\rm c}\right|^{-\gamma},
\end{eqnarray}
where $\alpha$, $\beta$, and $\gamma$ are critical exponents.
Each critical exponent does not have an independent value and $\alpha+2\beta+\gamma=2$ called the Rushbrooke relation is satisfied.
Although the relations given by Eq.~(\ref{RTeq:criticalexponents}) are behavior in equilibrium state, a similar relation exists in nonequilibrium process.
An order parameter which describes the second-order phase transition reaches to the equilibrium value with exponential decay except at the critical temperature.
However, the order parameter reaches to the equilibrium value with a power-law decay at the critical temperature.
The relaxation time $\tau$ relates to the correlation length $\xi$. 
When a second-order phase transition takes place, the relaxation time diverges in the thermodynamic limit, which is represented as
\begin{eqnarray}
 \tau \sim \xi^z,
\end{eqnarray}
where $z$ is called the dynamical critical exponent.
A typical example of system in which a second-order phase transition occurs is the ferromagnetic Ising model given by Eq.~(\ref{RTeq:IsingHam}).
The ferromagnetic Ising model is the simplest model and exhibits order-disorder transition with spontaneous symmetry breaking.

The critical slowing down relates to the domain-wall problem called the Kibble-Zurek phenomena\cite{RTKibble-1976,RTZurek-1985}.
The performance of annealing methods has been studied in terms of the Kibble-Zurek mechanism\cite{RTCaneva-2007,RTBiroli-2010,RTSuzuki-2011}.
Since the cost function of optimization problems with discrete variables can be represented by Hamiltonian of discrete spin systems, a second-order phase transition sometimes occurs as well as a first-order phase transition.
Then, in order to make annealing methods more efficient, we should avoid the second-order phase transition point in the protocol of annealing method.

\section{Potts model with invisible states}

In order to change phase transition nature with fixing a symmetry which breaks at the transition point, a new discrete spin model called Potts model with invisible states was proposed\cite{RTTamura-2010}.
In this section, we explain phase transition behavior of the model.
In condensed matter physics and materials science, when we find a phase transition with discrete symmetry breaking in real materials or complicated theoretical models, we often analyze the phase transition nature using the Potts model\cite{RTWu-1982}.
The Potts model is a cornerstone of discrete spin models in statistical physics.
In fact, the analysis using the Potts model succeeded in many cases\cite{RTWeger-1973,RTSzabo-1975,RTMukamel-1976,RTBarbara-1978,RTKim-1975,RTAharony-1977,RTBretz-1977,RTDomany-1977,RTBerker-1978,RTDomany-1978,RTDas-1979,RTDomany-1979,RTTejwani-1980,RTPark-1980,RTGouyet-1980a,RTGouyet-1980b,RTBlankschtein-1980a,RTBlankschtein-1980b,RTBlankschtein-1981,RTDomany-1982}.

Suppose we consider a phase transition with $q$-fold symmetry breaking in $d$ dimension.
In order to investigate the phase transition, we often refer the phase transition nature of ferromagnetic $q$-state Potts model on $d$-dimensional lattice.
The Hamiltonian of the model is given by
\begin{eqnarray}
 \label{RTeq:Potts}
 {\cal H} = -J\sum_{\langle i,j \rangle} \delta_{\sigma_i,\sigma_j},
  \qquad
  \sigma_i = 1, \cdots, q,
  \quad
  (q \in {\mathbb N}),
\end{eqnarray}
where the sum is over pairs of nearest-neighbor spins on $d$-dimensional lattice.
The Potts model is a generalized Ising model since the model is equivalent to the Ising model when $q=2$.
The ferromagnetic Potts model exhibits a phase transition at finite temperature in $d$ dimension ($d \ge 2$). 
Phase transition nature depends on the number of states $q$ and the spacial dimension $d$.
For example, in two dimension, when $q \le 4$, a second-order phase transition occurs whereas a first-order phase transition occurs when $q>4$.
In both cases, $q$-fold symmetry breaks at the phase transition point.
Not only the order of phase transition but also the critical phenomena were investigated.
In many cases, nature of phase transition observed in experiment and obtained in complicated theoretical model can be explained by the ferromagnetic Potts model.
Recently, however, phase transitions where the behavior is different from the ferromagnetic Potts model.
For instance, a first-order phase transition with threefold symmetry breaking was found in two-dimensional frustrated systems\cite{RTTamura-2008,RTOkumura-2010,RTTamura-2011}, although the three-state ferromagnetic Potts model in two dimension exhibits a second-order phase transition with threefold symmetry breaking.
As shown above, in the ferromagnetic Potts model, when the number of states $q$ and the spatial dimension $d$ are given, the order of the phase transition is determined.
Then, unconventional phase transitions such as the abovementioned examples cannot be represented by the ferromagnetic Potts model.

In order to overcome the fact, a generalized Potts model called Potts model with invisible states was proposed\cite{RTTamura-2010}.
The Hamiltonian is given by
\begin{eqnarray}
 \label{RTeq:Pottsinv_1}
 {\cal H} = -J \sum_{\langle i,j \rangle} \delta_{s_i,s_j} \sum_{\alpha=1}^q \delta_{s_i,\alpha},
  \qquad
  s_i = 1, \cdots, q+r,
  \quad 
(q+r \in {\mathbb N}).
\end{eqnarray}
If and only if $1 \le s_i = s_j \le q$, interaction between the $i$-th and $j$-th sites works.
Then, the state $s_i(\ge q+1)$ is regarded as redundant states.
Here the redundant states are called invisible states.
In the ground state, all spins have the same value from $1$ to $q$ and $q$-fold symmetry is broken.
When $r=0$, the model represented by Eq.~(\ref{RTeq:Pottsinv_1}) is equivalent to the standard Potts model given by Eq.~(\ref{RTeq:Potts}).
Thus, the model is a straightforward generalization of the Potts model.

In order to clarify the effect of invisible states, let us show another representation of the Potts model with invisible states when $r\ge 1$:
\begin{eqnarray}
 \label{RTeq:Pottsinv_2}
 {\cal H} = - J \sum_{\langle i,j \rangle} \delta_{\sigma_i,\sigma_j} \sum_{\alpha=1}^q \delta_{\sigma_i,\alpha}
  - T\ln r \sum_{i}\delta_{\sigma_i,0},
  \qquad
  \sigma_i = 0, 1, \cdots, q.
\end{eqnarray}
It should be noted that the Hamiltonian given by Eq.~(\ref{RTeq:Pottsinv_2}) is the same as that given by Eq.~(\ref{RTeq:Pottsinv_1}) at each temperature.
This fact can be confirmed by comparing partition functions of both models.
Note that the spin $s_i$ in the Hamiltonian given by Eq.~(\ref{RTeq:Pottsinv_1}) takes from $1$ to $q+r$ whereas the spin $\sigma_i$ in the Hamiltonian given by Eq.~(\ref{RTeq:Pottsinv_2}) takes from $0$ to $q$.
The invisible states are labeled by $\sigma_i = 0$ in Eq.~(\ref{RTeq:Pottsinv_2}).
The second term in the Hamiltonian given by Eq.~(\ref{RTeq:Pottsinv_2}) represents temperature-dependent chemical potential of invisible states, which is similar with the Wajnflasz-Pick model given by Eq.~(\ref{RTeq:WPt}).

Here we consider two-spin systems of the Potts model with invisible states given by Eq.~(\ref{RTeq:Pottsinv_1}) comparing with the standard ferromagnetic Potts model given by Eq.~(\ref{RTeq:Potts}).
In the standard ferromagnetic Potts model, the ground-state energy is $-J$ and the energy of excited states is $0$, which is the same as the Potts model with invisible states.
In both models, the number of ground states is $q$.
However the number of excited states in each model is different.
The number of excited states in the standard ferromagnetic Potts model is $q^2-q$ whereas that in the Potts model with invisible states is $q^2-q+2qr+r^2$.
Thus, the number of excited states increases due to existence of invisible states.
The increase of the number of excited states affects nature of phase transition.

Next we explain nature of phase transition in the Potts model with invisible states.
In Refs.~74, 98, and 99,~
the authors investigated phase transition behavior of the Potts model with invisible states in two dimension for $q \le 4$ and large $r$.
If there is no invisible states ($r=0$), a second-order phase transition occurs in the model for $q\le 4$.
The authors calculated temperature dependences of specific heat and order parameter which detects the $q$-fold symmetry breaking by Monte Carlo simulations.
As the temperature decreases, the order parameter becomes non-zero value at the temperature where the specific heat has the maximum value.
These behaviors suggest an existence of phase transition.
In order to determine the order of the phase transition, probability distribution of internal energy at the temperature where the specific heat has the maximum value was calculated.
The bimodal distribution was observed, which is a characteristic behavior of the first-order phase transition.
Moreover, the finite-size scaling analysis of the first-order phase transition was performed.
In the finite-size scaling analysis, the authors found that the latent heat remains in the thermodynamic limit.
From the above results, a first-order phase transition occurs in the Potts model with invisible states in two dimension for large $r$ even when $q \le 4$.
In addition, the authors confirmed that as $r$ increases, the transition temperature decreases but the latent heat increases.

The results obtained by Monte Carlo simulations suggest that the invisible states play a role to change phase transition nature.
In Refs.~74, 98, and 99,~
to confirm the fact, the authors also studied the phase transition of the Potts model with invisible states by the Bragg-Williams approximation which is a kind of mean-field analysis.
The Bragg-Williams approximation of the standard ferromagnetic Potts model given by Eq.~(\ref{RTeq:Potts}) concludes that a second-order phase transition occurs when $q=2$ whereas a first-order phase transition occurs when $q \ge 3$\cite{RTKihara-1954}.
Here we explain the Bragg-Williams approximation of the Potts model with invisible states.
For convenience, we use the representation of Hamiltonian given by Eq.~(\ref{RTeq:Pottsinv_2}).
Let $x_\alpha$ be the fraction of the $\alpha$-th state ($0 \le \alpha \le q$).
Obviously, $\sum_{\alpha=0}^q x_\alpha=1$ is satisfied.
Here $\alpha=0$ indicates the invisible state.
Since now we consider the case that the $q$-fold symmetry breaks at the transition point, one of $q$-states is selected in the ferromagnetically ordered phase.
The label of the selected state is set to $\alpha=1$.
Then the fractions are given by
\begin{eqnarray}
 &&x_0 = t,\\
  &&x_1 = \frac{1}{q}\left( 1 - t\right)\left[ 1 + \left( q-1\right)s\right],\\
  &&x_\alpha = \frac{1}{q}\left( 1-t\right)\left( 1 - s\right), \qquad (2\le \alpha\le q),
\end{eqnarray}
where $0 \le s,t \le 1$.
Then the internal energy $E^{\rm BW}$ and the entropy $S^{\rm BW}$ in the Bragg-Williams approximation are expressed as
\begin{eqnarray}
 \nonumber
 E^{\rm BW}(s,t) &&= - \frac{zJ}{2} \sum_{\alpha=1}^q x_\alpha^2 - x_0 T \ln r \\
 &&= -\frac{zJ(1-t)^2}{2q} \left[ \left( q - 1\right)s^2 + 1\right]- tT\ln r,\\
 \nonumber
  S^{\rm BW}(s,t) &&= - \sum_{\alpha=0}^q x_\alpha \ln x_\alpha\\
 &&= - t \ln t - \left( 1 - t\right)
\nonumber
  \left[
   \frac{1+(q-1)s}{q} \ln \frac{1+(q-1)s}{1-s} + \ln \frac{(1-t)(1-s)}{q}
  \right].\\
\end{eqnarray}
Then, the free energy is given by
\begin{eqnarray}
 F^{\rm BW}(s,t) = E^{\rm BW}(s,t) - TS^{\rm BW}(s,t).
\end{eqnarray}
By analyzing the free energy, we can obtain the transition temperature and latent heat.
As mentioned above, the Bragg-Williams approximation concludes that when there are no invisible states, a second-order phase transition occurs for $q=2$ and a first-order phase transition occurs for $q\ge 3$.
Then we first focus on the case of $q=2$.
When $(q,r)=(2,1),(2,2)$, and (2,3), a second-order phase transition with twofold symmetry breaking occurs.
In contrast, when $q=2$ and $r\ge 4$, a first-order phase transition occurs. 
In addition, the authors confirmed that when $q \ge 3$, a first-order phase transition occurs regardless of $r$.
Furthermore, as $r$ increases, the transition temperature decreases but the latent heat increases.
Then, the authors in Refs.~74, 98, and 99~
concluded that the invisible states play a role to change to a first-order phase transition from a second-order phase transition.
On other words, invisible states enlarge the latent heat and prevent the ordering.
After the authors proposed the model, the phase transition nature of the model on various lattices was investigated by analytical calculations\cite{RT-Enter2011,RT-Enter2012,RT-Mori2012,RT-Johnston2013,RT-Ananikian2013}.

Finally, we consider the structure of interaction in the Potts model with invisible states.
We show another representation of the Potts model with invisible states.
In the representation, the interaction tensor is used\cite{RTTamura-2012}.
We first consider the standard ferromagnetic Potts model.
Let $\vec{S}_i$ be a $q$-dimensional binary vector.
$\vec{S}_i$ represents the microscopic state in the $i$-th site.
Only one of elements in the vector is unity whereas the other elements are zero.
The position of unity indicates the state, {\it e.g.}, $\vec{S}_i={}^{\rm T}(0,1,0,\cdots,0)$ means that the state of the $i$-th spin is the second state.
Here the symbol ${\rm T}$ is the transpose of vector.
By using the vector representation, the Hamiltonian of standard ferromagnetic Potts model is given by
\begin{eqnarray}
 {\cal H} = - J \sum_{\langle i,j \rangle} \delta_{s_i,s_j}
  = - \sum_{\langle i,j \rangle} {}^{\rm T}\vec{S}_i \hat{J} \vec{S}_j, \qquad s_i = 1, \cdots, q,
\end{eqnarray}
where $\hat{J}$ is a $q\times q$ diagonal matrix:
\begin{eqnarray}
 \hat{J}= {\rm diag}(J,J,\cdots,J).
\end{eqnarray}
In a similar way, we can represent the Hamiltonian of Potts model with invisible states.
The Hamiltonian using the vector representation is given by
\begin{eqnarray}
 {\cal H} = -J \sum_{\langle i,j \rangle} \delta_{t_i,t_j} \sum_{\alpha=1}^q \delta_{t_i,\alpha}
  = -\sum_{\langle i,j \rangle} {}^{\rm T}\vec{T}_i \hat{J} \vec{T}_j, \qquad t_i = 1, \cdots, q+r,
\end{eqnarray}
where $\vec{T}_i$ is a $(q+r)$-dimensional binary vector and $\hat{J}$ is a $(q+r)\times (q+r)$ diagonal matrix:
\begin{eqnarray}
 \hat{J} = {\rm diag}(\underbrace{J,\cdots,J}_{q},\underbrace{0,\cdots,0}_{r}).
\end{eqnarray}
From a viewpoint of interaction structure, the results obtained in previous studies can be summarized as follows.
The order of phase transition can be changed by just expanding the space of the microscopic state.
The invisible states correspond to zero elements in the interaction tensor.
Some unconventional phase transitions found in two-dimensional frustrated systems can be represented by the Potts model with invisible states, which is an important progress in statistical physics and condensed matter physics.
Unfortunately, the method to change the nature of phase transition is not efficient for annealing methods since the order of phase transition is only changed in this method.
However, the method explained in this section is just a simple extension of the Potts model.
There are many remaining degrees of freedom, {\it e.g.}, off-diagonal elements in the interaction tensor.
Then, we believe that we can avoid a phase transition by employing a similar strategy.

\section{Conclusion and future perspective}

In this paper, we reviewed a method to change nature of phase transition toward annealing methods.
The annealing methods such as the simulated annealing and quantum annealing are regarded as an efficient general technique to solve optimization problems widely.
In Sec.~2, as an example of optimization problems, we introduced the traveling salesman problem. 
The traveling salesman problem can be represented by the Ising model.
In this way, most optimization problems with discrete variables can be represented by Hamiltonian of discrete spin systems.
Then, we can use generic algorithms proposed in terms of physics -- annealing method, to solve optimization problems.

In the simulated annealing, we introduce the temperature into optimization problems and gradually decrease the temperature.
On the other hand, in the quantum annealing, the quantum field such as a transverse field in the Ising model is introduced, and the quantum field is decreased.
The best solution of optimization problems can be definitely obtained when we decrease the temperature or quantum field slow enough.
In Sec.~3, mechanisms of the simulated annealing and quantum annealing were explained by using the Ising model on a square lattice as an example.

Although the annealing methods are versatile methods for optimization problems, we face with difficulties which come from phase transition in the annealing procedure.
We explained the difficulties when the first-order phase transition or second-order phase transition occur in the optimization problems in Sec.~4.
Then, in order to improve annealing method more efficient, we should avoid the phase transition in the annealing methods.
In Sec.~5, we showed a way to change nature of phase transition in the Potts model by introducing a new type of fluctuation called invisible state.
The invisible state is redundant state, and the ground state in the Potts model with invisible states does not change that in the standard ferromagnetic Potts model.
The authors in Refs.~74, 98, and 99~
concluded that the invisible states play a role to change to a first-order phase transition from a second-order phase transition.
Then, the method using the invisible states does not succeed to avoid the phase transition which induces the difficulty to obtain the best solution of optimization problems.
However, study on phase transition nature in the Potts model with invisible states is just getting started. 
It is an important issue to explore inherent properties of invisible states. 
In addition, extensions of the Potts model with invisible states are interesting, which was explained in the end of Sec. 5. 
Moreover, in view of optimization problems, investigation of the effect of invisible states in spin systems with inhomogeneous interactions is significant.

A way to change nature of phase transition using invisible states is easy to implement for optimization problems as well as the temperature and quantum field which are used in typical annealing methods.
Then, we strongly believe that underlying concept presented in this paper will be useful annealing methods to solve optimization problems.

\section*{Acknowledgements}
The authors are also grateful to Naoki Kawashima, Jie Lou, Yoshiki Matsuda, Seiji Miyashita, Takashi Mori, Yohsuke Murase, Taro Nakada, Masayuki Ohzeki, Per Arne Rikvold, Takafumi Suzuki, Yusuke Tomita, and Eric Vincent for their valuable comments. 
R.T. is partially supported by Grand-in-Aid for Scientific Research (C) (25420698) and National Institute for Materials Science (NIMS).
S.T. is partially supported by Grand-in-Aid for JSPS Fellows (23-7601).
The computations in the present work were performed on computers at the Supercomputer Center, Institute for Solid State Physics, University of Tokyo. 

\appendix{Monte Carlo method}

In Sec.~\ref{RTsec:SA}, we demonstrated the simulated annealing using the Monte Carlo method.
In this appendix, we show how to implement the Monte Carlo method.
Suppose we consider the Ising model with inhomogeneous interactions.
The Hamiltonian is given by
\begin{eqnarray}
 {\cal H} = -\sum_{\langle i,j \rangle} J_{ij} \sigma_i^z \sigma_j^z, \qquad (\sigma_i^z = \pm 1).
\end{eqnarray}
The procedure of Monte Carlo method is as follows:
\begin{description}
 \item[Step 1] We prepare an initial state.
 \item[Step 2] We choose a spin randomly.
 \item[Step 3] We calculate the local energy at the chosen site $i$.
	    The local energy is defined by
	    \begin{eqnarray}
	     h_i^{\rm (eff)}:= \sum_{j \,({\rm n.n.\,of}\, i)} J_{ij} \sigma_j^z,
	    \end{eqnarray}
	    where the summation is over the nearest-neighbor sites of the $i$-th site. 
	    Note that the Hamiltonian can be represented using the local energy:
            \begin{eqnarray}
            {\cal H} = -\frac{1}{2} \sum_i h_i^{\rm (eff)} \sigma_i^z.
            \end{eqnarray}
 \item[Step 4] We flip the chosen spin according to probability by some way.
	    In general, the probability can be calculated by the local energy, which will be explained later.
 \item[Step 5] We continue the procedure from Step 2 to Step 4.
\end{description}
There are two famous decision rules of the probability.
One is called the heat-bath method which is given by
\begin{eqnarray}
 p_{\rm HB}(\sigma_i^z \to -\sigma_i^z) = \frac{{\rm e}^{-\beta h_i^{\rm (eff)}\sigma_i^z}}{2 \cosh(\beta h_i^{\rm (eff)})}.
\end{eqnarray}
The other is called the Metropolis method which is given by
\begin{eqnarray}
 p_{\rm M}(\sigma_i^z \to -\sigma_i^z) = 
  \begin{cases}
   1 & (h_i^{\rm (eff)}\sigma_i^z < 0) \\
   {\rm e}^{-2\beta h_i^{\rm (eff)}\sigma_i^z} & (h_i^{\rm (eff)}\sigma_i^z \ge 0)
  \end{cases}.
\end{eqnarray}
Both of them satisfy the detailed balance condition. 
However, the detailed balance condition is just a sufficient condition for stochastic process toward equilibrium state. 
Then, a decision rule of the probability without detailed balance condition was proposed\cite{RT-Suwa2010,RT-Suwa2012,RT-Suwa2012a}.
Using the method, we can obtain the stable state efficiently.
Recently, a mechanism of the method has been studied in terms of nonequilibrium statistical physics\cite{RTShi-2012,RTOhzeki-2013,RTIchiki-2013,RTSakai-2013}.
In the simulated annealing, we decrease the temperature during the procedure from Step 2 to Step 5.

\appendix{Real-time dynamics by Schr\"odinger equation}

In Sec.~\ref{RTsec:QA}, we demonstrated the quantum annealing based on real-time dynamics.
In this appendix, we explain how to calculate real-time dynamics.
We first consider time-independent Hamiltonian.
The Schr\"odinger equation is given by
\begin{eqnarray}
 i \frac{\partial}{\partial t} \ket{\psi(t)} = \hat{{\cal H}} \ket{\psi(t)},
\end{eqnarray}
where the Planck constant $\hbar$ is set to unity.
The time evolution of wave function is expressed as 
\begin{eqnarray}
 \ket{\psi(t)} = {\rm e}^{-i\hat{\cal H}t}\ket{\psi(t=0)}=:\hat{U}(t)\ket{\psi(t=0)},
\end{eqnarray}
where $\hat{U}(t)$ is the time-evolution operator.
For time-independent Hamiltonians, we can immediately obtain the wavefunction at time $t$ if we assign the time $t$ and the initial wave function $\ket{\psi(t=0)}$.
In order to compute the time-evolution operator, the Hamiltonian should be diagonalized.
Let $\hat{\cal U}$ be unitary matrix which diagonalizes the Hamiltonian $\hat{\cal H}$.
Then, 
\begin{eqnarray}
 \hat{\cal H}_{\rm d} = \hat{\cal U}^\dagger \hat{\cal H} \hat{\cal U} = {\rm diag}(\epsilon_1,\cdots,\epsilon_{\cal D}),
\end{eqnarray}
where ${\cal D}$ is the number of microscopic states.
For $S=1/2$ spin system with $N$ sites, ${\cal D}=2^N$.
By using the unitary matrix $\hat{\cal U}$, the time-evolution operator is given by
\begin{eqnarray}
 U(t) = {\rm e}^{- i\hat{\cal H}t} = \hat{\cal U} {\rm e}^{-i \hat{\cal H}_{\rm d}t}\hat{\cal U}^\dagger.
\end{eqnarray}
Since the matrix $\hat{\cal H}_{\rm d}$ is a diagonal matrix, the matrix exponential is tractable:
\begin{eqnarray}
{\rm e}^{-i {\hat{\cal H}}_{\rm d} t} = {\rm diag}({\rm e}^{-i \epsilon_1 t}, \cdots, {\rm e}^{-i \epsilon_{\cal D} t}).
\end{eqnarray}

Next we consider the case that the Hamiltonian depends on time.
In this case, the Schr\"odinger equation is given by
\begin{eqnarray}
 i \frac{\partial}{\partial t}\ket{\psi(t)} = \hat{\cal H}(t)\ket{\psi(t)}.
\end{eqnarray}
The time evolution of wave function is formally described as
\begin{eqnarray}
 \label{RTeq:time-dependent-wv}
 \ket{\psi(t)} = \hat{\cal T} \exp 
  \left[
   - i \int_0^t {\rm d}t' \, \hat{\cal H}(t')
  \right]
  \ket{\psi(t=0)},
\end{eqnarray}
where $\hat{\cal T}$ is the time-ordered product of operators.
In the quantum annealing, we introduce a quantum field and decrease gradually the quantum field.
Then, we can obtain the time evolution of wave function by calculating Eq.~(\ref{RTeq:time-dependent-wv}).

\bibliographystyle{ws-procs9x6}

\end{document}